\documentclass[11pt]{article}
\usepackage[round]{natbib}
\usepackage{url}
\usepackage{hyperref}

\hypersetup{
bookmarks=true,     
colorlinks=true,       
linkcolor=red,          
citecolor=blue,        
filecolor=magenta,  
urlcolor=cyan          
}

\usepackage{authblk}

\usepackage{amssymb}
\usepackage{amsmath}

\usepackage{fancyhdr} 
\title{Long-term Productivity for Long-term Impact\footnote{Collegeville
    Workshop on Scientific Software Whitepapers, 2020,
    \url{https://collegeville.github.io/CW20/WorkshopResources/WhitePapers/WhitePaperList.html}}}

\author{Spencer Smith: \href{mailto:smiths@mcmaster.ca}{smiths@mcmaster.ca}}
\author{Jacques Carette: \href{mailto:carette@mcmaster.ca}{carette@mcmaster.ca}}
\affil{Computing and Software Department, McMaster
  University, Ontario, Canada}

\begin{document}

\maketitle

\begin{abstract}

  \noindent \textbf{Abstract:} \emph{We present a new conceptual definition of
    `productivity' for sustainably developing research software.  Existing
    definitions are flawed as they are short-term biased, thus devaluing
    long-term impact, which we consider to be the principal goal. Taking
    a long-term view of productivity helps fix that problem. We view the
    outputs of the development process as knowledge and user satisfaction.
    User satisfaction is used as a proxy for effective quality.  The
    explicit emphasis on all knowledge produced, rather than just the
    operationalizable knowledge (code) implies that human-reusable knowledge,
    i.e. documentation, should also be greatly valued when producing
    research software.}

\end{abstract}


%


%

\medskip

Research software can be a critical component of high-impact scientific and
engineering projects.  Some of that software is meant to be used just once (such
as when analyzing one-off data), but more often research software is meant to provide
useful tools to be re-used by the community.  While even one-off scripts should
be of decent quality and public, for the sake of reproducible science, here we
want to focus on the long-lived, impactful tools.

Such tools are meant to be long-lived, but a quick online search will find many
dead and abandoned projects.  Why is that?  In part, this is caused by the
short-term incentives inherent in the development environment (students leaving
upon completing their degree, the quest for new funding,
completion of the computations needed for a publication).  Meeting these goals can
be done even in the presence of sloppy software engineering practices.  However,
sloppy practices generate mountains of ``technical debt'' that are anathema to the
longevity and sustainability of long-lived impactful research software.

The focus on the short-term is a systemic problem, right down to the antiquated
definition of `productivity' that is often used (implicitly or explicitly). To
better understand how to balance short-term goals and long-term impact, we need
to revisit the very definition of `productivity' as it pertains to knowledge
work, especially when software is involved. This is an especially difficult
topic that has perplexed many for decades.  Here, we recommend Drucker's
foundational article ``Knowledge-Worker Productivity: The Biggest Challenge''
\citep{Drucker1999} as a lucid introduction to the problem.

To put it succinctly, we aim to provide a conceptual definition for `productivity' that
can be meaningfully applied in a research software context.

More precisely, we are \emph{re}defining productivity, as we found the
existing definitions inadequate (Section~\ref{Sec_Background}).
Sustainability, or the long-term development of useful software done with
a reasonably amount of effort, is not taken into account. In
Section~\ref{Sec_Criteria} we lay out the criteria that underly our
(re)definition. Finally in Section~\ref{Sec_Redefinition}, we explain our
redefinition of productivity.

\section{Defining Productivity: a Brief History} \label{Sec_Background}

The first scientific study of productivity of workers, in a manufacturing
context, was conducted by \citet{Taylor1911} over $100$ years ago.  He
first focused on examining and understanding the tasks that needed to be done,
eliminating those that were not needed, and then optimizing the remaining ones.
It was ground-breaking, and enabled decades of productivity improvements. It
also showed that labourer productivity can be increased by using incentives and
investing in training. His insights have been relabelled many times over the
decades, but not fundamentally changed.

The standard definition of productivity, such as given by
\href{https://www.bls.gov/lpc/} {U.S. Bureau of Labor Statistics}, is to divide
the amount of goods and services produced by the inputs used in their production
(usually a combination of labour and capital).  The
\href{http://www.businessdictionary.com/definition/productivity.html} {Business
  Dictionary} augments this slightly, by computing productivity as the ratio of
average output per period and the total costs incurred or resources (capital,
energy, material, personnel) consumed in that period.  In an agricultural
context, productivity measures the amount produced by a target group (country,
industry, sector, farm or almost any target group) given a set of resources and
inputs \citep{FAO2017}.  The details differ, but these equations all come down
to defining productivity as outputs produced per unit of input.  The time period
of interest is relatively short; it is the time between starting and finishing
production.

The inputs and outputs of ``production'' for knowledge-based work are much
harder to pin down. Labour hours spent on-task is certainly inadequate.  And
what about the necessary hours spent on administration, or learning new skills?
What about ``day dreaming''?  Both research and anecdotal evidence suggest that
some of the most impactful ideas result from a wandering mind
\citep{Sundheim2020}.  What are the outputs?  In the words of
\citet{Drucker1999}, we need to know \emph{what is the task?}  Is the only work
that matters related to the executable code, or should we count test cases,
design documentation, meeting minutes, etc?  When do we define a task as
complete?  When the code first compiles, or its first released, or its
$20^{\text{th}}$ release?  Merely releasing software is less than half the work,
as the quality of the results is at least as important as the quantity
\citep{Drucker1999}.

In software engineering \citet{Boehm1987} defines productivity as above: outputs
divided by inputs. For him, the inputs are comprised of labor, computers,
supplies, and other support facilities and equipment, accounted in present-value
dollars. What specifically to count as input is left open for each project; \citet{Boehm1987}
suggests each project's manager decide whether to include costs such as
requirements analysis, documentation, project management and secretarial support.
Worse still, for outputs, \citet{Boehm1987} mentions several options including
Delivered Source Instructions (DSI), code complexity metrics and function
points.  All of these have been thoroughly debunked as meaningful outputs. Agile
projects, for a while, measured productivity via story
points \citep{Infopulse2018}, but that too has been soundly rejected; academic
papers on such negative results are hard to find, but cogent explanations by
software engineers abound; we recommend \citet{Nortal}.  Our position is that the
problems with productivity measures are caused by too close an adherence to the
manual worker model proposed by \citet{Taylor1911}.

\citet{Drucker1999} helpfully gives six major factors underlying knowledge
worker (KW) productivity:
\begin{enumerate}
\item Clearly asking \emph{What is the task?}
\item Workers must have \emph{autonomy} on how to perform their tasks. They must
  largely manage themselves.
\item Continuous improvement has to be part of the work and responsibility of
  KW.
\item Continuous learning, and continuous teaching (i.e. knowledge sharing) are
  crucial and expected.
\item Productivity should be weighted at least as heavily on quality as on
  quantity.
\item Productivity accounting requires that KW are viewed as assets, not costs.
\end{enumerate}

What \citet{Drucker1999} is implicitly saying is
\begin{itemize}
\item \emph{Knowledge} is both an input and an output to knowledge work,
\item The software is \emph{not} the task.
\end{itemize}

\section{Improving the Definition of Productivity} \label{Sec_Criteria}

Existing definitions of productivity largely ignore effects that arise
because of long term concerns, such as sustainability.  Furthermore, their views
of what are inputs and outputs are too simplistic regarding knowledge work.
 
\subsection{Long-Term View}

Sustainable development ``... meets the needs of the present without
compromising the ability of future generations to meet their own needs''
\citep{Brundtland1987}.  Meeting the needs of the present entails developing
research software that meets its requirements, both in terms of quality
and functionality; meeting the needs of the future means software that is
maintainable, reusable, and can be used in reproducible research. Properly taking
the future into account requires a future-viewing definition of productivity.

This entails:
\begin{enumerate}
\item that quantity cannot be the only measure of productivity, quality is at
  least as important~\citep{Drucker1999}.  If current work does not survive into
  the future, then it is not part of long-term productivity.  If current work
  adds technical debt, it should be considered negative productivity.
\item that code is not the only important artifact.  On large projects, turnover
  and training are substantial issues.  Documentation becomes a crucial means by
  which project knowledge does not simply disappear.
\item that productivity should be based on \emph{outcomes}, not artifacts
  produced.  For much research software, this outcomes are a combination of
  impact, utility over time and reach.
\end{enumerate}

On the last point, an apt analogy is perhaps tenure: the goal is to grant tenure
to faculty members that have produced quality work (work of lasting value) over
a long period of time, as judged by their respective communities, in the hopes
that they will continue to do into the future.

\citet{Infopulse2018} emphasizes outcome based productivity by suggesting that
business success is the ultimate metric of productivity.  The key question to
ask: ``Is your customer happy?''. For research software, this is better termed
as \emph{user satisfaction}, as this encompasses both open source and commercial
views of software.

\subsection{Outputs: Knowledge and Satisfaction}

The creation of research software is much more than just encoding
previous knowledge as executable knowledge (aka code).
Code is merely an ends to a means: to encode certain knowledge in
executable form, for the benefit of many. But code is a terrible means
of \emph{knowledge transmission}. Furthermore, the creation of research
software usually involves creating new knowledge.
\citet{Drucker1999} implicitly says the same thing when he links the
productivity of KWs to continuous teaching, since documentation is a form of
teaching, or knowledge transfer.  

What about ``user satisfaction''? Certainly the numbers of users, number of
citations, number of forks of a repository, and the number of ``stars'',
weighted by the potential number of users, are all indicators that users
find utility in the software.  A proper satisfaction measure would likely
also involve considering both known issues and a survey of existing users.

By redefining output to encompass the quality and quantity of knowledge
produced, previous measures' flaws can be eliminated.  For instance, counting
lines of codes becomes explainably bad.  Counting lines of code discourages
using an external library, refactoring to make code shorter, or even spending
time writing a requirements specification. However, all of these are activities
that increase the effective knowledge delivered over time to users, at a
lesser total cost.

As the knowledge produced will be used by different kinds of users (such
as internal developers and external users), it is important to weigh
the satisfaction of each class separately.

Specific measures would take us too far afield, but we can point to some
useful evidence.  A structured process backed by tools can
work very well: \citet{SmithEtAl2018_StatSoft} shows that the quality of
statistical software for psychology is considerably higher for CRAN
(Comprehensive R Archive Network) based software than others. The main
difference is that CRAN mandates that some content be present in any
contribution.  This is backed by automated verification tools.

\subsection{Inputs: Labour and Knowledge}

Research software is created by skilled, knowledgeable people, who ought to be
able to efficiently integrate both their internal (tacit) knowledge of the task
to be done with external, relevant knowledge. The latter can come in multiple
forms, both internal to the project (manuals, design documents, theory
manuals, etc.) and external (textbooks, papers, etc.)  We cannot really
measure this by days worked: some studies show \citep[p.\ 468]{GhezziEtAl2003}
that the most important productivity factor, by a wide margin, is the
capability of the personnel. In other words, a junior engineer may struggle
for days to accomplish a task that a senior colleague could complete in minutes.
The importance of knowledge for KWs is why \citep{Drucker1999} emphasizes the
importance of continuous learning.

Making knowledge an explicit input should dispel the common
misconception \citep[p.\ 469]{GhezziEtAl2003} that software developers are
interchangeable.  Knowledge disparity between workers is one reason why a
software development project cannot be sped up by simply assigning more
developers to the task \citep{Brooks1995}.  Another reason is communication
overhead for sharing the required knowledge.

Knowledge appears as both an input and an output because knowledge is the
feedback in the software production loop.  Part of the knowledge produced while
developing one version of the software will be used as input for the next
iterations (often far into the future) of the software.  Moreover, knowledge
produced by other projects may also be relevant, and should be incorporated.

However, we need to differentiate between actual inputs and effective
inputs. The \emph{actual inputs} are the number of hours worked by all
workers associated to the project, regardless of task or skill. The
\emph{effective inputs} are those hours that are spent on actually generating
knowledge.

\section{Productivity Redefined} \label{Sec_Redefinition}

It is still too early to give a directly measurable definition of productivity.
We thus settle on a definition that can at least allow us to reason about
productivity, one that encompasses more of the relevant factors.  Scientific
computing (the heart of much research software) is our model: its definition of
(forward) error requires knowing the, usually unknown, true answer.

As a starting point, here is our conceptual formula for productivity:

$$ I = \int_{0}^{T} H(t)\ dt $$
$$ O = \int_{0}^{T} \sum_{c \in C} S_c(t) K_c(t)\ dt $$
$$P = O / I$$ 

\noindent where $P$ is productivity, $0$ is the time the project started, $T$ is the
time \emph{in the future} where we want to take stock, $H$ is the total number
of hours available by all personnel, $C$ represents different classes of users
(external as well as internal), $S$ is satisfaction and $K$ is \emph{practical
knowledge}.  Thus productivity is measured in ``satisfying reusable
knowledge per hour.''

An obvious refinement would be to split the time period into two, $\left[0 ,
\text{now}\right]$ and $\left[\text{now}, T\right]$, and then the quantities in
the future integrals could be modified to be over \emph{expectations} of the
quantities in question. This would more clearly account for the effect of
the ``crystal ball'' that must be used to predict what will be practical
knowledge and what will increase user satisfaction.

\section{Conclusion}

%

We have presented a conceptual definition that emphasizes that productivity for
knowledge based work, like research software, only makes sense when considered
over the long term.  If our goal is sustainable software, then we need a
definition of productivity that will drive developers to value long-term
considerations.  In particular, this means explicitly considering knowledge as
an output.  One practical consequence would be an increased emphasis on
internal documentation meant to capture crucial internal knowledge.  By shifting
the time-frame by which productivity is computed, we have exposed the inherent
worth of traditionally under-valued tasks --- and thus hope to shift misplaced
attitudes/priorities.

Considering long-term productivity requires minimizing total effort of all
people involved over the total lifetime of a project.  This does not mean that
we are advocating BDUF (Big Design Up Front)!  By explicitly integrating over a
long period, we accumulate the incremental value produced by successive
iterations.  BDUF is not only unrealistic, it also produces lower value.  It is
possible to incrementally design, build and document research software, and even
fake it as if a rational (up front appearing, omniscient) process were followed
\citep{SmithEtAl2019_arXiv}.

An important component of our definition is an emphasis on
user satisfaction.  Quality is crucial, especially in knowledge work like
software, but hard to directly measure. But we cannot decouple quality
from usefulness: a bug-free software that no one uses is not particularly
valuable.  Thus we can legitimately use a mixed proxy: Adoption by the
user community. This is not foolproof: if poor quality software is the only
one that exists, and it still provides a useful-enough service, it will
be used. Luckily, time and/or competition can solve that problem.

Our definition can already be used to better understand various tradeoffs
in long-term software development practices. Naturally, deeper investigations
will inevitably leads to refinements. Ideally, we'll also find robust means
of approximating our measure of productivity. This should allow us to
more accurately determine the impact of different processes, methodologies and
tools on the development of sustainable, impactful research software.


\end{document}